# Software Reuse in Medical Database for Cardiac Patients using Pearson Family Equations

M. Bhanu Sridhar
Dept. of CSE,
Raghu Engineering College,
Visakhapatnam.

Y. Srinivas
Dept. of IT,
GITAM University,
Visakhapatnam.

M.H.M. Krishna Prasad,
Dept. of IT,
JNTUK Campus,
Vizianagaram.

## ABSTRACT

Software reuse is a subfield of software engineering that is used to adopt the existing software for similar purposes. Reuse Metrics determine the extent to which an existing software component is reused in new software with an objective to minimize the errors and cost of the new project. In this paper, medical database related to cardiology is considered. The Pearson Type-I Distribution is used to calculate the probability density function (pdf) and thereby utilizing it for clustering the data. Further, coupling methodology is used to bring out the similarity of the new patient data by comparing it with the existing data. By this, the concerned treatment to be followed for the new patient is deduced by comparing with that of the previous patients' case history. The metrics proposed by Chidamber and Kemerer are utilized for this purpose. This model will be useful for the medical field through software, particularly in remote areas.

## 1. INTRODUCTION

Software Engineering is an eminent field of Computer Science that can be used to organize, assess, analyse, realize and implement the software with a useful methodology. New ideas have been coming up since the word of Software Engineering was coined by Peter Naur and Brian Randell at the NATO Science Committee Conference in 1968 [1]. Different models have evolved for its expression; numerous metrics for measuring the relative field have been brought out and different sub-fields of the concept have also taken their place.

One of the prominent sub-fields of Software Engineering is the concept of Software Reuse. Software Reuse is a means to improve the practice of Software Engineering by using the existing software artefacts during the construction of new software systems [2]. Reuse aims at bringing down the effort put into the work, increasing the productivity and results in large-scale development of efficient software. Apparently the productivity increases since, the developers does not need to bring up the new concept from the scratch; they can make use of the available components and artefacts. Since the existing software is already 'proved', the new software that makes a reuse of it can be reliable. Software Reuse is not limited to any part of Software Engineering – it can be used in various levels of the software development life cycle. Requirements, analysis, design; coding, testing and even maintenance can be considered as the relevant fields where Software Reuse is extremely useful for all purposes. Software Reuse has been advocated as a mechanism to reduce productivity time, increase the production scale and decrease defect density [2]. However, little importance has been given for the concept of software reuse in the medical domain.

Software Reuse in medical field is ever useful as reusing the medicines themselves.In this paper, the concept of software reusability is used with respect to the medical domain, in particular for cardiology. Cardiology is a medical specialty dealing with human heart disorders. This field includes diagnosis and treatment of disorders like heart defects, heart failure and other heart diseases. According to World Health Organization, India has the highest number of coronary heart disease deaths in the world [3]. This can be deduced not only due to lack of resources but also due to concentration of resources at cities and towns. By usage of Internet and cardiology database component reuse, the Para-medics, can deduce the medicines or methods to be used for the patients at remote places to temporarily put them out of danger. From the reuse of available data, the required medicines may also be deduced and can be proposed to the patients [4].

The data collected for a patient with a problem of cardiology, which can be reused in the context of other patients, to deduce if the concerned patients are near, inside or outside the clutches of the disease. Similarly, the medicines used previously in a similar condition will be very handy when a patient with a similar condition is detected. Reuse is counted as vital in medical field since previous information is very handy in deducing a patient's current health position and save the precious life [5].

In this article, the database of the cardiac patients from archives [6] is considered for the application of this methodology. The paper is organised as follows: Section-1 of the paper deals with introduction; Pearson family of probability distributions is presented in Section-2; in Section-3, the considered dataset is discussed and presented; Section-4 deals with coupling and metrics, where the methodologies and results are presented and discussed. Finally the conclusion is presented in Section-5.

In the future work, which is now at an advanced research stage, should be very useful to aid the ailing patients and can become an important part in the general convention of the Doctors.

## 2. THE PEARSON FAMILY OF PROBABILITY DISTRIBUTIONS

While dealing with medical domain, it is necessary to identify the disease accurately. This is because the same type of diseases with similar features, when occurred to patients suffering from similar symptoms, in remote areas can be treated in the similar fashions or at least minimum lifesaving drugs can be prescribed. Traditional clustering algorithms such as K-Means may suit the purpose, but the efficiency of K-means is based on the value of 'K', the number of clusters. Falsifying the K-value leads to misclassification and this necessitated the development of generalized systems of probability distributions. Karl Pearson obtained a family of distributions, known as the Pearson distributions, which would represent satisfactorily all the practical situations.

The probability density function of Pearson type-1 distribution is proposed by Karl Pearson [7]. Pearson's examples comprise survival data, which are usually asymmetric.

The particular cases of Pearson family of distributions include Gaussian, beta, and Student's t-distributions. The Probability Density function of theType-1, Pearson family [8] is given by Eqn. (1)

$$\frac{df(x)}{dx} = \left\{ \frac{b+x}{a_0 + a_1 x + a_2 x^2} \right\} \tag{1}$$

The variable $X$ denotes the grey level intensity value of the echocardiographic speckle of a cardiac patient and $f(x)$ represents the probability density function (pdf). The $b$, $a_0$, $a_1$, and $a_2$ are parameters of the distribution. The parameters can be evaluated using the moments given by the following equation.

$$b = -a_1 = \frac{0.5 \mu_3 (3 \mu_2^2 + \mu_4)}{9 \mu_2^3 - 5 \mu_2 \mu_4 + 6 \mu_3^2}$$

$$a_0 = \frac{0.5 \mu_2 (4 \mu_2 \mu_4 - 3 \mu_3^2)}{9 \mu_2^3 - 5 \mu_2 \mu_4 + 6 \mu_3^2}$$

$$a_2 = \frac{0.5 (6 \mu_3^2 - 2 \mu_2 \mu_4 + 3 \mu_3^2)}{9 \mu_2^3 - 5 \mu_2 \mu_4 + 6 \mu_3^2} \tag{2}$$

In the Pearson family, the selection parameter is evaluated from the first four moments; the skewness and the kurtosis of the model are given by

Skewness, $S_k = \mu_3 / \mu_2^{3/2}$         (3)

Kurtosis, $K_u = \mu_4 / \mu_2^2$         (4)

Using the above skewness and kurtosis values, the value of k is evaluated by the formula

$$\kappa = \frac{S_k^2 (K_u + 3)^2}{4 (4 K_u - 3 S_k^2)(2 K_u - 3 S_k^2 - 6)} \tag{5}$$

The Pearson family of distributions are basically categorised into three major types Type I, Type IV and Type VI are defined for $\kappa < 0$, $0 < \kappa < 1$ and $\kappa > 1$, respectively. The pdf of Type I distribution (k<0) is taken into consideration which is given by

$$f_1(x) = A_0 (1 + \frac{x - m_0}{c_1})^{g_1} (1 - \frac{x - m_0}{c_2})^{g_1}, -c_1 + m_0 < x < c_2 + m_0$$

(6)

where

$$A_0 = \frac{g_1^{g_1} g_2^{g_2} \Gamma(g_1 + g_2 + 2)}{(c_1 + c_2)(g_1 + g_2)^{g_1 + g_2} \Gamma(g_1 + 1) \Gamma(g_2 + 1)}$$

$$g_{2,1} = 0.5h - 1 \pm sign(\mu_3)(0.5h(h+2)) \frac{S_k}{\sqrt{S_k^2 (h+2)^2 + 16h + 16}}$$

$$c_1 = (\frac{g_1}{g_2}) \frac{0.5 \sqrt{\mu_2 \{ S_k^2 (h+2)^2 + 16h + 16 \}}}{1 + \frac{g_1}{g_2}}$$

$$c_2 = 0.5 \sqrt{\mu_2 \{ S_k^2 (h+2)^2 + 16h + 16 \}} - c_1$$

$$h = \frac{6 K_u - 6 S_k^2 - 6}{6 + 3 S_k^2 - 2 K_u} \tag{7}$$

$$m_0 = \mu_1 - 0.5 \frac{\mu_3 (h+2)}{\mu_2 (h-2)}$$

These models are used to assess the symptoms of the patients and utilizing this model for developing a methodology that can be reused.

## 3. DATASET

The dataset considered here is that of patients who have symptoms of cardiac problems. The cardiac problems are briefly discussed below.

## 3.1 Cardiac Problems

The heart is a myogenic (cell-related) muscular organ with a circulatory system (including all vertebrates), that is responsible for pumping blood throughout the blood vessels by repeated, rhythmic contradictions [4]. Among the problems related to heart, the major problem is cardiac arrest, which is the cessation of normal blood circulation due to failure of the heart to contract effectively. It should be effectively realised that cardiac arrest is different from a heart attack where blood supply is interrupted to a part of the heart which may/may not lead to the patient's death.

The patients who approach a doctor can be classified into three categories taking into consideration results of different tests conducted with the existing symptoms. The properties taken into consideration are Atherosclerosis (due to Cholesterol), Myocardial Infarction (heart attack), different medical signs like blood cell count and skin rashness, various symptoms like head ache and body pain, and other facts like Diabetes, Triglyceride, Migraine and so on. The three categories are:

(a) Normal: A patient can be declared 'normal' when no signs or symptoms of a cardiovascular/coronary disease are found within the results of various tests conducted. The general factors considered are the blood pressure (BP), sugar level in blood, results of Electrocardiography (ECG), Cholesterol level, Triglyceride, and other sensations.

(b) Pro-Cardiac: Pro-cardiac category keeps the account of those patients who are suspected to have some signs and/or symptoms of heart-problems. These can be observed from the BP tests slightly exceeding the normal levels, sugar levels in blood also rising, ECG suspecting (though not deducing) problems in future and some signs and symptoms like light chest pain, high cholesterol, severe headaches often turning up etc. do surface.

(c) Cardiac: A cardiac is surely in the range of trouble – prone to abnormal BP conditions, having severe pain the chest region, burning sensations, sweating, pain along the left arm and finally having already had a light heart attack. A cardiac must be immediately taken into consideration for regular treatment with constant observation of all concerned positions in and around the heart and those that affect the heart.

The predominant features considered in the paper for the cardiology database are: blood pressure (BP), heartbeat (HB), pulse rate (PR), ECG (normal/abnormal), pain in the left shoulder region, sweating, nausea/vomiting, overweight, chest pain and breathlessness. Eleven core symptoms are considered while preparing the patient data dissimilarity matrix.

After the discussion of all the classification parts, it should also be noted that effective medical data of the concerned patient should be readily available for the Doctors which should be regularly updated. This data forms the pillar of the patient's classification level, severity level and the chance of saving his/her life [9]. An attempt is made in this paper, by bringing into picture the reuse of data, to correctly judge the patient's position and hence suggest the correct drugs.

Depending on the latest reports of the patient, he can be recognised to be prone by a known disease and further classified into what level of that disease he is in. A dissimilarity matrix is constructed with the readings from the data and identifying the most leading symptoms. The various readings considered are categorized into the above mentioned three groups for a patient who had hit trouble with cardiology. A database is formulated from the realistic data obtained from medical patients from the data referred in [6].

For the testing purpose a database of twenty patients has been used, with the symptoms of cardiac problems, taking into consideration eleven core features/symptoms. If a symptom does exist it has been represented by using a unique value. Following this procedure for all the inputs, a dissimilarity matrix [10] is obtained and this matrix is to be categorized. To classify a patient, the dissimilarity matrix is formulated and classified by using the clustering technique in the context of coupling.

## 4. METHODOLOGY AND RESULTS
## 4.1 Coupling

A component is more likely to be fault-free, if its functionality is distributed appropriately among its subcomponents. Appropriate distribution of function brings about coupling and cohesion. Coupling is the extent to which various subcomponents interact with each other [11]. If components are highly interdependent, changes on one function may affect the other functions also. Hence loose coupling is generally desirable and is a sign of well-structured computer system.

Coupling can be classified into different types based on their inter-dependability, as, content coupling, common coupling, external coupling, control coupling, data coupling and so on. It should be noted that in object-oriented coupling is a measure of the strength of association between the modules and is classified as interaction coupling, component coupling and inheritance coupling.

Each of these coupling dimensions deduces that the behaviour of a class C is inter-dependent on the behaviour of another class $C^1$ [12] such that a change in C is reflected in $C^1$ as well. The degree of coupling can be described as how complex this information can be. Here, C is an existing patient and $C^1$ is the new patient.

The flow of measurements in coupling proceeds from low level high level coupling. Low coupling is desirable since in terms of software maintenance and reuse, the software quality apparently increases. Upgrading and replacing is easier to plan in this context since inter-dependence is not much a factor to consider. On the other end, high coupling is a complex relationship making reuse or maintenance much problematic.

**Table 1: Symptoms (→) of the patients**

| Patient ID (↓) | BP | Heart beat (HB) | Pulse Rate (PR) | ECG | Left Shoulder pain | Swea ting | Vomi ting | Over Weigh t | Chest Pain | Breath lessne ss | Obesity |
|---|---|---|---|---|---|---|---|---|---|---|---|
| P1 | 0 | 2 | 0 | 4 | 0 | 6 | 7 | 0 | 0 | 0 | 0 |
| P2 | 0 | 0 | 0 | 4 | 5 | 6 | 7 | 0 | 0 | 0 | 0 |
| P3 | 0 | 0 | 0 | 4 | 0 | 0 | 0 | 0 | 9 | 0 | 0 |
| P4 | 0 | 0 | 0 | 0 | 0 | 0 | 0 | 0 | 0 | 0 | 0 |
| P5 | 0 | 2 | 0 | 4 | 0 | 6 | 7 | 0 | 0 | 0 | 0 |
| P6 | 0 | 0 | 3 | 4 | 5 | 6 | 7 | 8 | 9 | 10 | 11 |
| P7 | 0 | 0 | 0 | 0 | 0 | 0 | 0 | 0 | 0 | 10 | 0 |
| P8 | 0 | 0 | 0 | 4 | 5 | 6 | 7 | 0 | 0 | 0 | 0 |
| P9 | 0 | 0 | 0 | 0 | 0 | 6 | 0 | 0 | 0 | 10 | 0 |
| P10 | 0 | 0 | 3 | 0 | 5 | 0 | 7 | 0 | 9 | 0 | 11 |
| P11 | 0 | 2 | 0 | 0 | 0 | 6 | 7 | 8 | 9 | 0 | 11 |
| P12 | 0 | 0 | 3 | 0 | 0 | 0 | 0 | 0 | 9 | 10 | 0 |
| P13 | 0 | 0 | 0 | 4 | 5 | 6 | 7 | 0 | 0 | 0 | 0 |
| P14 | 0 | 2 | 0 | 0 | 5 | 6 | 7 | 8 | 0 | 0 | 0 |
| P15 | 0 | 2 | 0 | 0 | 5 | 6 | 0 | 0 | 0 | 0 | 11 |
| P16 | 0 | 2 | 0 | 4 | 0 | 6 | 7 | 0 | 0 | 0 | 0 |
| P17 | 0 | 2 | 0 | 0 | 5 | 6 | 0 | 0 | 0 | 0 | 11 |
| P18 | 0 | 0 | 0 | 4 | 0 | 0 | 7 | 8 | 9 | 10 | 0 |
| P19 | 0 | 2 | 0 | 0 | 5 | 0 | 0 | 0 | 0 | 0 | 11 |
| P20 | 0 | 2 | 0 | 4 | 0 | 6 | 7 | 0 | 0 | 0 | 0 |

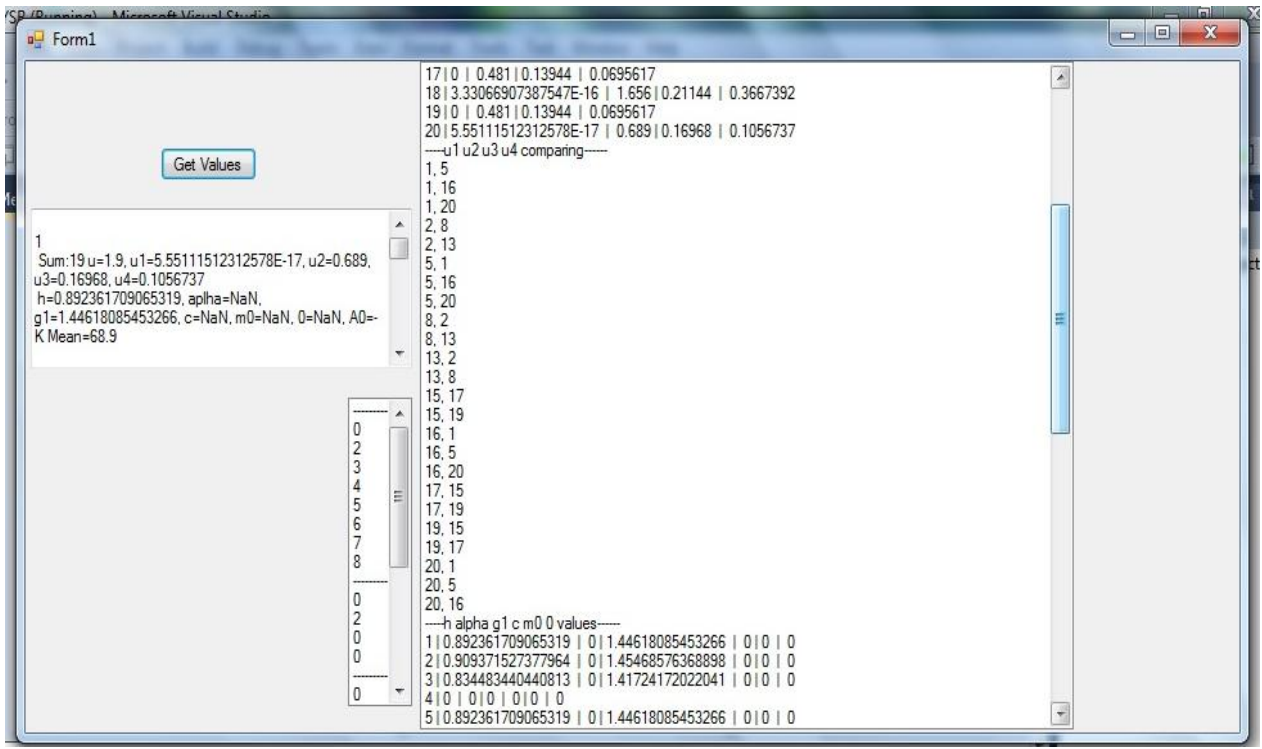

**Figure 1: The results of coupling**

The data considered is that of twenty patients of cardiology (main component) in which represent eleven important symptoms (sub-components) are represented. By adapting to coupling, the patients are classified in terms of their present sub-components i.e., number of patients with BP, no. of patients with chest pain and so on. By the usage of Pearson Type-I distribution equations, the probability density function (pdf) is calculated for each patient.

When the data of a new patient is obtained, the maximum likelihood function is calculated first. Basing on this the cluster to which this patient belongs is identified and thereby the categorization is carried out. By this it is concluded that the new patient is most similar to a particular old patient. Hence the medicines already used for the old patient can be prescribed for the new patient also or at least a first-aid drug is suggested.

From the above model, the considered dataset is clustered into three groups based on the symptoms. The clustering results show that the patients P1, P5, P16, and P20 exhibit the similar symptoms and hence they fall into the same cluster, patients P2, P8 and P13 exhibit the same symptoms and hence can be grouped into one cluster. Patients P15, P17, P19 are clustered into a group. Once the clustering process is completed, it facilitates the concerned persons at the primary health centres in remote areas to take up the necessary follow-up action for the patients having similar case-histories. The same drugs that are administered can be reused to the patients with similar symptoms so that the first aid can be provided and thereby decreasing the mortality rate.

Some of the coupling results with symptoms under consideration are given below:

The dataset under consideration is that of 20 patients with the test results of 11 symptom tests of cardiology.

1. Statistics of single symptom coupling:

**Table 2: Single Symptom Coupling**

| Sl. No. | Symptom | Count | Patient IDs |
|---------|---------|-------|-------------|
| 1. | 3 | 3 | 6,10,12 |
| 2. | 8 | 4 | 6,11,14,18 |
| 3. | 10 | 5 | 6,7,9,12,18 |
| 4. | 9 | 6 | 3,6,10,11,12,18 |
| 5. | 11 | 6 | 6,10,11,15,17,19 |
| 6. | 2 | 8 | 1,5,11,14,15,16,17,20 |
| 7. | 5 | 9 | 2,6,8,10,13,14,15,17,19 |
| 8. | 4 | 10 | 1,2,3,5,6,8,13,16,18,20 |
| 9. | 7 | 12 | 1,2,5,6,8,10,11,13,14,16,18,20 |
| 10. | 6 | 14 | 1,2,5,6,8,9,11,13,14,15,16,17,19,20 |

2. Statistics of first four symptoms coupling:

**Table 3: Four symptoms coupling**

| Sl. No. | Count | Patient IDs |
|---------|-------|-------------|
| 1. | 1 | 6 |
| 2. | 2 | 10,12 |
| 3. | 3 | 4,7,9 |
| 4. | 4 | 1,5,16,20 |
| 5. | 5 | 2,3,8,13,18 |
| 6. | 5 | 11,14,15,17,19 |

3. Statistics of all symptoms coupling:

**Table 4: Total Symptom Coupling**

| Sl. No. | Count | Patient IDs |
|---------|-------|-------------|
| 1. | 3 | 2,8,13 |
| 2. | 3 | 15,17,19 |
| 3. | 4 | 1,5,16,20 |

## 4.2 Reuse Metrics

In order to measure the amount of reuse attained in these models, the metrics suite proposed by Chidamber and Kemerer [13] is taken into consideration. The suite proposed for object oriented design and is used in this paper. It is important to note that [13] defined coupling as 'two objects are coupled if changes on one object affect the other object'. Further, two classes are coupled if changes in once class use properties of the other class.

A metric of this suite, the metric-4, is adopted which is concerned about 'coupling between objects' (CBO). Here it is to be once again noted that excessive coupling is negative to the basic component design and prevents reuse. The CBO for a class is defined to be the number of other classes to which it is coupled. In this context a patient exhibiting different symptoms is considered to be a class and the symptoms are the objects.

The dataset considered above is used in this context to apply the CBO metric to view the results.

**Table 5: Coupling count of each class (patient)**

| Patient ID | Associated with | Coupling Count |
|------------|-----------------|----------------|
| 1 | 5,16,20 | 3 |
| 2 | 8,13 | 2 |
| 5 | 1,16,20 | 3 |
| 8 | 2,13 | 2 |
| 13 | 2,8 | 2 |
| 15 | 17,19 | 2 |
| 16 | 1,5,20 | 3 |
| 17 | 15,19 | 2 |
| 19 | 15,17 | 2 |
| 20 | 1,5,16 | 3 |
| 3,4,6,7,9,10,11,12,14,18 | NONE | 10 |

**Table 6: Application of CBO Metric**

| No. of Patients | CBO Metric Value |
|-----------------|------------------|
| 4 | 3 |
| 6 | 2 |
| 10 | 0 |

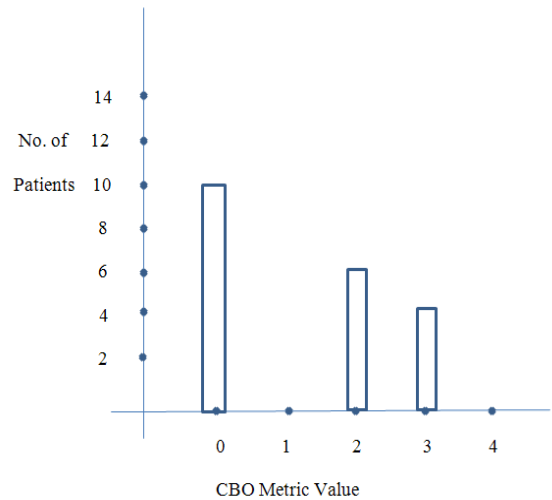

**Figure 2: Histograms for CBO Metric**

Figure 2 depicts the results obtained concerning the patients and symptoms, using the CBO metric.

After all the process the patients are classified into different groups or clusters. Since 10 patients are not associated with much of the symptoms, they may be classified as 'normal'. The next group with a value of 6 associated with considerable number of symptoms are can be placed in the cluster of 'pro-cardiac'. Finally, the last 4 patients are surely of 'cardiac' group since they are associated with the highest no. of symptoms.

If the data of a new patient is now provided, he is easily classified into one of the concerned clusters and is provided the same medicines as had been done to the previous patients.

## 5. CONCLUSION

In this paper, a new dimension based on software reuse for the medical domain is presented, in particular, to cardiology. A database of 20 patients is considered and is categorised into 3 categories depending upon the health conditions. The readings for these categories are obtained from the doctors and prominent websites, and are used for testing the reusability. The dissimilarity matrix is generated where each symptom is indexed. By the usage of Pearson Type-1 set of equations, the probability density function values for each patient are generated and are grouped into different clusters. Next, the coupling methodology is brought into picture and the different clusters, based on their coupling values, are found out. Further, by the usage of metric-4, from the object oriented metrics suite of Chidamber and Kemerer the number of patients (classes) with common symptoms (objects) are obtained which confirms the group division.

By this developed methodology, one can notice that the number of patients with no common symptoms is 10 and can classify them as 'normal'. Next the 'pro-cardiac' group is obtained as patients in the group with CBO metric value of 2. Finally the all-dangerous 'cardiac' group is obtained from the remaining CBO metric value of 3. At this juncture, armed with full data calculations and conclusions, if a new patient's data is forwarded, one can easily place him in a cluster, and treatment can be started as it was done for the previous patients of that cluster.

This article aims mainly for patients at remote areas, where only primary healthcare centres are available. These centres have blood testing and ECG as the premium sources of reasoning the patient's symptoms and the disease, in particular. If this model is made available for these remote area centres, the paramedics can just supply data to the model, find out the similarity of this new data with one of the old patients' data and provide the all-important first-aid so as to save the life of the ailing patient.